%% file: main.tex
\documentclass{sig-alternate}

\usepackage{amsmath}
\usepackage{amsfonts}
\usepackage{amssymb}
\usepackage{mathtools}
\usepackage{url}
\usepackage{cite}
\usepackage{xspace}
\usepackage{graphicx}
\usepackage{subfigure}
\usepackage{color}
\usepackage{stmaryrd}
\usepackage{multirow}
\usepackage{algorithm}
\usepackage{listings}
\usepackage{algpseudocode}
\usepackage{multicol}
\usepackage{booktabs}
\usepackage{subfig}

\makeatletter
\def\@copyrightspace{\relax}
\makeatother

\newif\ifspace\spacefalse
\graphicspath{{figs/}}

\newtheorem{definition}{Definition}

\renewcommand{\qed}{\ensuremath{~\blacksquare}}

\graphicspath{{./}}

\newcommand{\alphabet}{\mathrm{\Sigma}}

\newcommand{\tru}{\top}
\newcommand{\fals}{\bot}
\newcommand{\LTL}{{\sc Ltl}\xspace}

\newcommand{\FLTL}{{\sc Fltl}\xspace}
\newcommand{\LTLfour}{{\sc Ltl}$_4$\xspace}
\newcommand{\Forall}{\mathbb{A}\xspace}
\newcommand{\Exists}{\mathbb{E}\xspace}

\newcommand{\F}{\mathbf{F}}
\newcommand{\G}{\mathbf{G}}
\newcommand{\X}{\mathbf{X}}
\newcommand{\U}{\mathbf{U}}
\newcommand{\TC}{{\top_c}}
\newcommand{\FC}{{\bot_c}}
\newcommand{\TT}{{\top_p}}
\newcommand{\FT}{{\bot_p}}

\newcommand{\Q}{\mathcal{Q}}

\newcommand{\Fistar}{\hat{\varphi}}
\newcommand{\BigQ}{\mathbb{Q}_\varphi}
\newcommand{\BigD}{\mathbb{D}_{\varphi,u}}
\newcommand{\D}{D_{\varphi}}

\newcommand{\monitor}{\mathcal{M}}

\newcommand{\LTLFOC}{{\sc Ltl}$_4-${\sc C}\xspace}

\newcommand\code[1]{\textsf{\small #1}}

\begin{document}
	
\input{abs}
 \input{intro}
 \input{problem}
 \input{monitor}

 \input{design}
 \input{experiments}
 \input{related}
 \input{conclusion}

\bibliographystyle{plain}
\bibliography{bibliography}

\end{document}

%% file: abs.tex
\title{Accelerated Runtime Verification of \\ LTL Specifications with Counting Semantics}


\numberofauthors{1}

\author{
%
%
\alignauthor
Ramy Medhat, Yogi Joshi, Borzoo Bonakdarpour, and Sebastian Fischmeister\\
       \affaddr{University of Waterloo, McMaster University, Canada}\\
       \email{\{rmedhat, y2joshi, sfischme\}@uwaterloo.ca, 
borzoo@McMaster.ca}\\
}
\maketitle

\begin{abstract}
	
Runtime verification is an effective automated method for specification-based 
offline testing and analysis as well as online monitoring of complex systems. 
The specification language is often a variant of regular expressions or a 
popular temporal logic, such as \LTL. This paper presents a novel and efficient 
parallel algorithm for verifying a more expressive version of 
\LTL specifications that incorporates counting semantics, where nested quantifiers can be subject to numerical constraints. Such constraints are useful in evaluating thresholds 
(e.g., expected uptime of a web server). The significance of this extension is 
that it enables us to reason about the correctness of a large class of systems, 
such as web servers, OS kernels, and network behavior, where properties are 
required to be instantiated for parameterized requests, kernel objects, network 
nodes, etc. Our algorithm uses the popular {\em MapReduce} architecture to split 
a program trace into variable-based clusters at run time. Each cluster is then 
mapped to its respective monitor instances, verified, and reduced collectively 
on a multi-core CPU or the GPU. Our algorithm is fully implemented and we report 
very encouraging experimental results, where the monitoring overhead is 
negligible on real-world data sets.

\end{abstract}

%

%% file: intro.tex
\vspace{-3mm}
\section{Introduction}

In this paper, we study runtime verification of properties specified in an 
extension of linear temporal logic (\LTL) that supports expression of counting 
semantics with  numerical constraints. Runtime verification (RV) is an automated
specification-based technique, where a {\em monitor} evaluates the correctness 
of a set of logical properties on a particular execution either on the fly 
(i.e., at run time) or based on log files. Runtime verification complements 
exhaustive approaches such as model checking and theorem proving and 
under-approximated methods such as testing. The addition of counting semantics 
to properties is of particular interest, as they can express parametric 
requirements on types of execution entities (e.g., processes and threads), user- 
and kernel-level events and objects (e.g., locks, files, sockets), web services 
(e.g., requests and responses), and network traffic. For example, the 
requirement `every open file should eventually be closed' specifies a rule for 
causal and temporal order of opening and closing individual objects which 
generalizes to {\em all} files. Such properties cannot be expressed using 
traditional RV frameworks, where the specification language is propositional 
\LTL or regular expressions.

In this paper, we extend the 4-valued semantics of \LTL (i.e, \LTLfour), 
designed for runtime verification~\cite{bls10-jlc} by adding counting semantics 
with  numerical constraints and propose an efficient parallel algorithm 
for their verification at run time. Inspired by the work 
in~\cite{libkin2004elements}, the syntax of our language (denoted \LTLFOC) 
extends \LTL syntax by the addition of {\em counting quantifiers}. That is, we 
introduce two quantifiers: the {\em instance counting quantifier} ($\Exists$) 
which allows expressing properties that reason about the number of satisfied or 
violated instances, and the {\em percentage counting quantifier} ($\Forall$) 
which allows reasoning about the percentage of satisfied or violated instances 
out of all instances in a trace. These quantifiers are subscripted with 
numerical constraints to express the conditions used to evaluate the count. For 
example, the following 
\LTLFOC formula:
$$\Forall_{\ge 0.95} \, s \, \text{:}\, \code{socket}(s) \Rightarrow 
\left( \G \,\code{receive}\left(s\right) \implies \textbf{F} \, 
\code{respond}\left(s\right) \right)$$
intends to express the property that `at least $95\%$ of open TCP/UDP sockets 
must eventually be closed'. also, the formula:
$$\Forall x : \code{user}(x) \Rightarrow \left(\Exists_{\le3} \,r:\code{rid}(r) 
\Rightarrow \left( \code{login} \wedge \code{unauthorized} \right) \right)$$
intends to capture the requirement that `for all users, there exist at most 
$3$ requests of type login that end with an unauthorized status'. The semantics of \LTLFOC is defined over six truth values:

\begin{itemize}
\item {\bf True} ($\tru$) denotes that the property is already permanently 
satisfied.

\item {\bf False} ($\fals$) denotes that the property is already permanently 
violated.

\item {\bf Currently true} ($\TC$) denotes that the current execution satisfies 
the quantifier constraint of the property, yet it is possible that an extension 
violates the constraint.

\item {\bf Currently false} ($\FC$) denotes that the current execution 
violates the quantifier constraint of the property, yet it is possible that an 
extension satisfies it.

\item {\bf Presumably true} ($\TT$) denotes that the current execution 
satisfies the inner \LTL property and the quantifier constraint of the 
property.

\item {\bf Presumably false} ($\FT$) denotes that the current execution 
violates the inner \LTL property and the quantifier constraint of the 
property.
\end{itemize}
We claim that these truth values provide us with informative verdicts about 
the status of different components of properties (i.e., quantifiers and their 
numerical constraints as well as the inner \LTL formula) at run time.

The second contribution of this paper is a divide-and-conquer-based online 
monitor generation technique for \linebreak \LTLFOC specifications. In fact, 
\LTLFOC monitors have to be generated at run time, otherwise, an enormous 
number of monitors (in the size of cross-product of domains of all variables) 
has to be created statically, which is clearly impractical. Our technique first 
synthesizes an \LTLfour monitor for the inner \LTL property of \LTLFOC 
properties pre-compile 
time using the technique in~\cite{bls10-jlc}. Then, based upon the values of 
variables observed at run time, submonitors are generated and merged to compute
the current truth value of a property for the current program trace.

Our third contribution is an algorithm that implements the above approach for 
verification of \LTLFOC properties at run time. This algorithm enjoys two levels 
of parallelism: the monitor (1) works in parallel with the program under 
inspection, and (2) evaluates properties in a parallel fashion as well. While 
the former ensures that the runtime monitor does not intervene with the normal 
operation of the program under inspection, the latter attempts to maximize the 
throughput of the monitor. The algorithm utilizes the popular {\em MapReduce} 
technique to (1) spawn submonitors that aim at evaluating subformulas using 
partial quantifier elimination, and (2) merge partial evaluations to compute 
the current truth value of properties.

Our parallel algorithm for verification of \LTLFOC properties is fully 
implemented on multi-core CPU and GPU technologies. We report rigorous 
experimental results by conducting three real-world independent case studies. 
The first case study is concerned with monitoring HTTP requests and responses 
on an Apache Web Server. The second case study attempts to monitor users 
uploading maximum chunk packets repeatedly to a personal cloud storage service 
based on a dataset for profiling DropBox traffic. The third case study monitors 
a network proxy cache to reduce the bandwidth usage of online video services, based on a YouTube request dataset. We 
present performance results comparing single-core CPU, multi-core CPU, and GPU 
implementations. Our results show that our GPU-based implementation provides an 
average speed up of $7$x when compared to single-core CPU, and $1.75$x when 
compared to multi-core CPU. The CPU utilization of the GPU-based implementation 
is negligible compared to multi-core CPU, freeing up the system to perform more 
computation. Thus, the GPU-based implementation manages to provide competitive 
speedup while maintaining a low CPU utilization, which are two goals that the 
CPU cannot achieve at the same time. Put it another way, the GPU-based 
implementation incurs minimal monitoring costs while maintaining a 
high throughput.

The rest of the paper is organized as follows. Section~\ref{sec:prob} describes 
the syntax and semantics of \LTLFOC. In Section~\ref{sec:monitor}, we explain 
our online monitoring approach, while Section~\ref{sec:design} presents our 
parallelization technique based on MapReduce. Experimental results are 
presented in Section~\ref{sec:exp}. Related work is discussed in 
Section~\ref{sec:related}. Finally, we make concluding remarks and discuss 
future work in Section~\ref{sec:concl}.

%% file: problem.tex
\vspace{-3mm}
\section{LTL with counting semantics} \label{sec:prob}

To introduce our logic, we first define a set of basic 
concepts.

\begin{definition}[Predicate]
Let $V = \{x_1,x_2,\dots, x_n\}$ be a set of variables with (possibly 
infinite) domains \linebreak $\mathcal{D}_1, \mathcal{D}_2,\dots, 
\mathcal{D}_n$, respectively. A 
{\em predicate} $p$ is a binary-valued function on the domains of variables in 
$V$ such that $$p: \mathcal{D}_1 \times \mathcal{D}_2 \times \cdots \times 
\mathcal{D}_n \, 
\rightarrow \, \{\text{true},\text{false}\}~\blacksquare$$
\end{definition}
The arity of a predicate is the number of variables it accepts. A predicate 
is {\em uninterpreted} if the domain of variables are not known concrete sets. 
For instance, $p(x_1,x_2)$ is an uninterpreted predicate, yet we can interpret 
it as (for instance) a binary function that checks whether or not $x_1$ is less 
than $x_2$ over natural numbers.

Let $\mathit{UP}$ be a finite set of uninterpreted predicates, and let 
$\Sigma=2^{\mathit{UP}}$ be the power set of $\mathit{UP}$. We call each element 
of $\Sigma$ an {\em event}.
\begin{definition}[Trace]
A {\em trace} $w=w_0w_1\cdots$ is a finite or infinite sequence of events; 
i.e, $w_i \in \Sigma$, for all $i \geq 0$.\qed
\end{definition}
We denote the set of all infinite traces by $\Sigma^{\omega}$ and the set of 
all finite traces by $\Sigma^{*}$.
A {\em program trace} is a sequence of events, where each event consists of 
{\em interpreted} predicates only. For instance, the following trace is a 
program trace:
$$w=\{\code{open}(1),\code{r},\code{anony})\}\, 
\{\code{open}(2),\code{rw},\code{user}(5)\} \, \cdots$$
where \code{open} and \code{user} are unary predicates and 
\code{r}, \code{anony}, and \code{rw} are 0-arity predicates. 
Predicate \code{open} is interpreted as opening a file, \code{r} is
interpreted as read-only permissions, \code{anony} is interpreted as an 
anonymous user, and so on.


\subsection{Syntax of LTL4-C}
\LTLFOC extends \LTLfour with two counting quantifiers: the instance counting quantifier ($\Exists$) and the percentage counting quantifier ($\Forall$). The semantics of these quantifiers are introduced in subsection~\ref{subsec:semantics}. The syntax of \LTLFOC is defined as follows:

\begin{definition}[\LTLFOC Syntax]
	\label{def:syntax}
	 \LTLFOC formulas \linebreak are defined using the following grammar: 
	\begin{equation} \nonumber
		\begin{split}
			\varphi \, \mathtt{::=} \,&\Forall_{\sim k} \;x:p(x) \Rightarrow \varphi \; \mid \; \Exists_{\sim l} \;x:p(x) \Rightarrow\varphi \; \mid \;  \psi \\
			\psi \, \mathtt{::=} \, & \top \; \mid \; 
p\left(x_1 \cdots x_n\right)\; \mid \; \neg \psi \;\mid \psi_1  \wedge \psi_2 
\; \mid \; \\
			& \X \,\psi \; \mid \; \psi_1\,\U\, \psi_2
		\end{split}
	\end{equation}
	\noindent where $\Forall$ is the percentage counting quantifier, $\Exists$ is the instance counting quantifier, $x$, $x_1\cdots x_n$ are variables with possibly 
infinite domains $\mathcal{D}, \mathcal{D}_1,\cdots \mathcal{D}_n$, $\sim \, 
\in \left\{ <,\le,>,\ge,= \right\}$, $k \, \mathtt{:} \, \mathbb{R} \in 
\left[0,1\right]$, $l \, \in \mathbb{Z^+}$, $\X$ is the next, and $\U$ is 
the until temporal operators.\qed
\end{definition}
If we omit the numerical 
constraint in $\Forall_{\sim k}$ (respectively, 
$\Exists_{\sim l}$), we mean $\Forall_{= 1}$ (respectively, $\Exists_{\ge 1}$). 
The syntax of \linebreak \LTLFOC forces constructing formulas, where a string 
of counting quantifiers is followed by a quantifier-free formula. We emphasize that 
$\Forall$ and $\Exists$ do not necessarily resemble standard first-order 
quantifiers $\forall$ and $\exists$. In fact, as we will explain $\neg 
\Forall$ and $\Exists$ are not generally equivalent.

Consider \LTLFOC property $\varphi = \Forall x : p(x) \Rightarrow 
\psi$, where the domain of $x$ is $\mathcal{D}$. This property denotes that for 
any possible valuation of the variable $x$ ($[x:=v]$), if $p(v)$ holds, then 
$\psi$ should hold. If $p(v)$ does not hold, then $p(v) \Rightarrow \psi$ 
trivially evaluates to true. This effectively means that the quantifier $\Forall 
x$ is in fact applied only over the following sub-domain: $$\{v \in \mathcal{D} 
\mid p(v)\} \subseteq \mathcal{D}$$

To give an intuition, consider the scenarios where file management anomalies 
can cause serious problems at run time (e.g., in NASA's Spirit Rover on Mars 
in 2004). For example, the following \LTLFOC property expresses ``at least half of the files that a process has previously opened must be closed'':
\begin{equation}
	\label{eq:example}
	\varphi_1 = \Forall_{\ge 50\%} \,f: \code{intrace}(f) \Rightarrow 
(\code{opened}(f) \,\U\, \code{close}(f))
\end{equation}
where $\code{intrace}$ denotes the fact that the concrete file appeared in 
any event in the trace.

\subsection{4-Valued LTL~\cite{bls10-jlc}}

First, we note that the syntax of \LTLfour can be easily obtained 
from Definition~\ref{def:syntax} by (1) removing the counting quantifier rules and (2) 
reducing the arity of predicates to 0 (i.e., predicates become atomic 
propositions).

\subsubsection{FLTL}
To introduce \LTLfour semantics, we first introduce Finite \LTL. Finite \LTL 
(\FLTL)~\cite{mp95} allows us to reason about finite traces for verifying 
properties at run time. The semantics of \FLTL is based on the truth values 
$\mathbb{B}_2=\{\tru,\fals\}$.

\vspace{-2mm}
\begin{definition} [\FLTL semantics]
Let $\varphi$ and $\psi$ be \LTL properties, and $u = u_0u_1 \cdots u_{n-1}$ 
be a finite trace.
	\begin{align*}
\left[u \models_{\text{\normalfont F}} \X\, \varphi\right] &= 
\begin{cases}
	[u_1 \models_{\text{\normalfont F}} \varphi] & \text{\normalfont if } u_1 \neq \epsilon \\
	\fals & \text{\normalfont otherwise}
\end{cases}\\ \\
\left[u \models_{\text{\normalfont F}} \varphi \,\U\, \psi\right] &= 
\begin{cases}
		\tru & \exists k \in [0,n-1]: [u_k \models_{\text{\normalfont F}} \psi] = \tru \;\;\wedge \\
	&\forall l \in [0,k) : [u^l \models_{\text{\normalfont F}} \varphi] = \tru\\
	\fals & \text{\normalfont otherwise}
\end{cases}
\end{align*}
where $\epsilon$ is the empty trace. The semantics of \FLTL for atomic 
propositions and Boolean combinations are identical to that of \LTL. \qed
\end{definition} 
Similar to standard \LTL, $\F p \equiv \tru \, \U \,p$ and $\G p \equiv \neg \F 
\neg p$.

\subsubsection{LTL4 Semantics}
\LTLfour is designed for runtime verification by producing more informative 
verdicts than \FLTL. The semantics of \LTLfour is defined based on values 
$\mathbb{B}_4=\{\tru,\TT,\FT,\fals\}$ ({\em true}, {\em presumably true}, {\em 
presumably false}, and {\em false} respectively). The semantics of \LTLfour is 
defined based on the semantics \LTL and \FLTL.

\begin{definition} [\LTLfour semantics]
\label{def:ltlfour}
Let $\varphi$ be an \LTLfour \linebreak property and $u$ be a finite prefix of a trace.
\begin{equation*}
\left[u \models_4 \varphi\right] = 
\begin{cases}
	\tru & \forall v \in \alphabet^\omega : uv \models \varphi\\
	\fals & \forall v \in \alphabet^\omega: uv \not \models \varphi\\
	\TT & [u \models_F \varphi] \, \wedge \, \exists v \in 
\alphabet^\omega: uv \not \models \varphi \\
	\FT & [u \not \models_F \varphi] \, \wedge \, \exists v 
\in \alphabet^\omega: uv \models \varphi\qed
\end{cases}
\end{equation*}
\end{definition}
In this definition, $\models$ denotes the satisfaction relation defined by standard \LTL semantics over infinite traces. Thus, an \LTLfour property evaluates to $\tru$ with respect to a finite trace 
$u$, if the property remains {\em permanently satisfied}, meaning that for all 
possible infinite continuations of the trace, the property will always be 
satisfied in \LTL. Likewise, a valuation of $\fals$ means that the property will 
be {\em permanently violated}. If the property evaluates to $\TT$, this denotes 
that currently the property is satisfied yet there exists a continuation that 
could violate it. Finally, value $\FT$ denotes that currently the property is 
violated yet there exists a continuation that could satisfy it.

\subsubsection{LTL4 Monitors}
\begin{figure}[t]
 \centering
  \includegraphics[width=0.5\columnwidth]{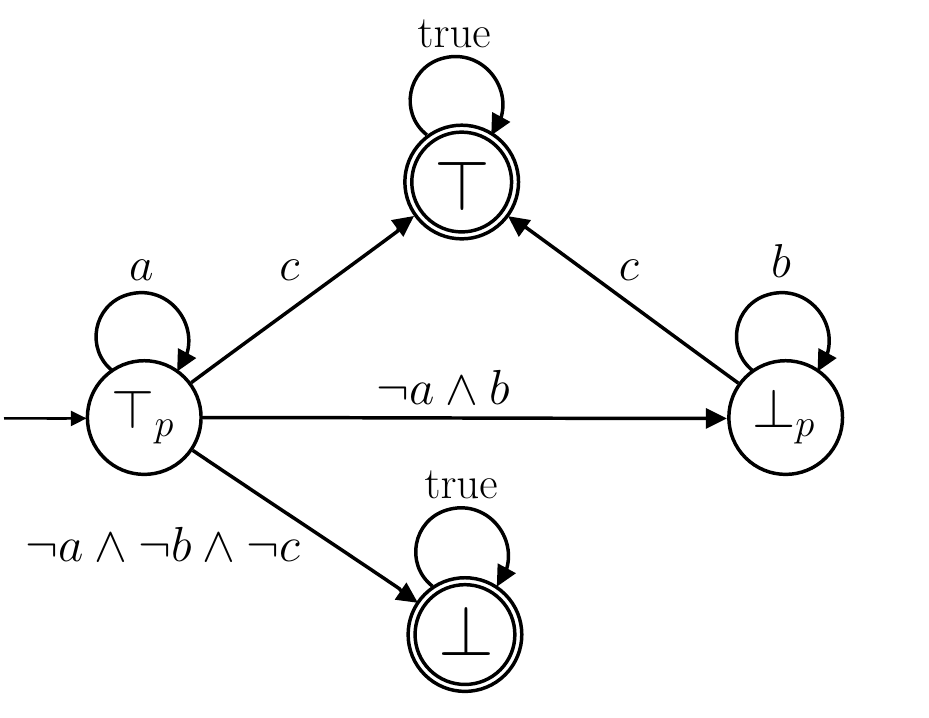}
 \caption{\LTLfour monitor for property $\varphi = \G a \; \vee \; (b \, \U \, 
c)$.}
 \label{fig:ltl4}
\vspace{-3mm}
\end{figure}

In \cite{bls10-jlc}, the authors introduce a method of synthesizing a 
{\em monitor}, as a deterministic finite state machine (FSM), for an \LTLfour 
property.
\begin{definition} [\LTLfour Monitor]
\label{def:monitor2}
Let $\varphi$ be an \LTLfour formula over $\alphabet$. The {\em 
monitor} $\monitor_{\varphi}$ of $\varphi$ is the unique FSM $(\Sigma, Q, q_0, 
\delta, \lambda)$, where $Q$ is a set of states, $q_0$ is the initial state, 
$\delta \subseteq Q \times \Sigma \times Q$ is the transition relation, and 
$\lambda: Q \rightarrow \mathbb{B}_4$ is a function such that:
\begin{equation*}
\left[u \models_4 \varphi\right] = \lambda(\delta(q_0, u)).~\blacksquare
\end{equation*}
\end{definition}
Thus, given an \LTLfour property $\varphi$ and a finite trace $u$, monitor 
$\monitor_{\varphi}$ is capable of producing a truth value in $\mathbb{B}_4$, 
which is equal to $[u \models_4 \varphi]$. For example, Figure~\ref{fig:ltl4} 
shows the monitor for property $\varphi = \G a \; \vee \; (b \, \U \, c)$. 
Observe that a monitor has two {\em trap} states 
(only an outgoing self loop), which map to 
truth values $\tru$ and $\fals$. They are trap states since these truth values 
imply permanent satisfaction (respectively, violation). Otherwise, states 
labeled by $\TT$ and $\FT$ can have outgoing transitions to other states.

\subsection{Truth Values of LTL4-C}

The objective of \LTLFOC is to verify the correctness of quantified properties 
at run time with respect to finite program traces. Such verification attempts to 
produce a sound verdict regardless of future continuations. 

We incorporate six truth values to define the semantics of \LTLFOC: 
$\mathbb{B}_6=\{\tru, \fals, \TC, \FC, \TT, \FT\}$; {\em true}, {\em false}, 
{\em currently true}, {\em currently false}, {\em presumably true}, {\em 
presumably false}, respectively. The values in $\mathbb{B}_6$ form a lattice ordered as 
follows: $\fals < \FC < \FT < \TT < \TC < \tru$. Given a finite trace $u$ and 
an \LTLFOC 
property $\varphi$, the informal description of evaluation of $u$ with respect 
to $\varphi$ is as follows: 
\begin{itemize}

\item {\bf True} ($\tru$) denotes that any infinite extension of $u$ 
satisfies $\varphi$. 

\item {\bf False} ($\fals$) denotes that any infinite extension of $u$ 
violates $\varphi$.  

\item {\bf Currently true} ($\TC$) denotes that currently $u$ 
satisfies the counting quantifier constraint of $\varphi$, yet it is possible that a 
suffix of $u$ violates the constraint. For instance, the valuation of
Property~\ref{eq:example} (i.e., $\varphi_1$) is $\TC$, 
if in a trace $u$, currently $50\%$ of files previously opened are 
closed. This is because (1) the inner \LTL property is permanently satisfied 
for at least $50\%$ of files previously opened, and (2) it is possible for a 
trace continuation to change this percentage to less than $50\%$ in the 
future (a trace in which enough new files are opened and not closed).

\item {\bf Currently false} ($\FC$) denotes that currently $u$ violates the 
quantifier constraint of $\varphi$, yet it is possible that a suffix of $u$ 
satisfies the constraint. For instance, the valuation 
of Property~\ref{eq:example} (i.e., $\varphi_1$) in a finite trace $u$ is 
$\FC$, if the number of files that were not successfully opened is 
currently greater than $50\%$. This could happen in the scenario where opening a file fails, possibly due to lack of permissions.  Analogous to $\TC$, the property is evaluated to $\FC$ because (1) the inner 
\LTL property is permanently satisfied for less than $50\%$ of files in the 
program trace, and 
(2) it is possible for a trace continuation to change this percentage to at 
least $50\%$ in the future.

\end{itemize}

Now let us consider modifying the property to support multiple open and close operations on the same file. For this purpose, we reformulate the property as follows:
\begin{equation}
	\label{eq:example2}
	\varphi_2 = \Forall_{\ge 50\%} \,f: \code{intrace}(f) \Rightarrow 
\left(\G \left(\code{opened}(f) \,\U\, \code{close}(f)\right)\right)
\end{equation}

\begin{itemize}
\item {\bf Presumably true} ($\TT$) extends the definition of {\em presumably 
true} in \LTLfour~\cite{bls10}, where $\TT$ denotes that $u$ satisfies the inner 
\LTL property and the counting quantifier constraint in $\varphi$, if the program 
terminates after execution of $u$. For example, Property~\ref{eq:example2} 
(i.e., $\varphi_2$) evaluates to $\TT$, if at least $50\%$ of the files in the 
program trace are closed. Closed files presumably satisfy the property, since 
they satisfy the $\G$ operator thus far, yet can potentially violate it if 
the file is opened a subsequent time without being closed. Note that this 
property can never evaluate to $\TC$, since no finite trace prefix can 
permanently satisfy the inner \LTL property. However, if the inner property can 
be permanently satisfied ($\tru$) and presumably satisfied ($\TT$), then the 
entire \LTLFOC property can potentially evaluate to $\TC$ if the numerical 
condition of the quantifier is satisfied. A property can evaluate to $\TT$ only 
if the conditions for $\TC$ are not met, since $\TC$ is higher up the partial 
order of $\mathbb{B}_6$.

\item {\bf Presumably false} ($\FT$) extends the definition of {\em
presumably false} in \LTLfour~\cite{bls10}, which denotes that $u$ 
presumably violates the quantifier constraint in $\varphi$. According to 
the Property~\ref{eq:example2}, this scenario will occur when the
number of files that are either closed or opened and not yet closed is at least 
$50\%$ of all files in the trace. Opened files presumably violate the inner 
property, since closing the file is required but has not yet occurred. This 
condition should not conflict with $\TT$ or $\TC$, since they 
precede $\FT$ in the partial order of $\mathbb{B}_6$ and thus $\FT$ only occurs 
if the conditions for $\TT$ and $\TC$ do not hold.


\end{itemize}


\subsection{Semantics of LTL4-C}
\label{subsec:semantics}
An \LTLFOC property essentially defines a set of traces, where each traces is a
sequences of events (i.e., sets of uninterpreted predicates). We define the 
semantics of \LTLFOC with respect to finite traces and present a method of 
utilizing these semantics for runtime verification. In the context of runtime 
verification, the objective is to ensure that a program trace (i.e., a 
sequence of sets of {\em interpreted} predicates) is in the set of traces that 
the property defines, given the interpretations of the property predicates 
within the program trace.  

To introduce the semantics of \LTLFOC, we examine counting quantifiers 
further. Since the syntax of \LTLFOC allows nesting of counting quantifiers, a 
canonical form of properties is as follows:
\begin{equation}
	\label{eq:genprop}
	\varphi = \BigQ  \; \psi
\end{equation}
where $\psi$ is an \LTL property and $\BigQ$ is a string of 
counting quantifiers
\begin{equation}
	\label{eq:bigq}
\BigQ=\Q_0\Q_1\cdots\Q_{n-1}
\end{equation}
such that each $\Q_i = \langle
Q_i,\sim_i,c_i,x_i,p_i\rangle$, $0 \leq i \leq n-1$, is a tuple encapsulating 
the counting quantifier information. That is, $Q_i \in \{\Forall,\Exists\}$, $\sim_i \in 
\left\{ <,\le,>,\ge,= \right\}$, $c_i$ is the constraint constant, $x_i$ is the 
bound variable, and $p_i$ is the predicate within the quantifier (see 
Definition~\ref{def:syntax}).

We presents semantics of \LTLFOC in a stepwise manner:
\begin{enumerate}
\item \textbf{Variable valuation.} First, we demonstrate how variable 
valuations are extracted from the trace and used to substitute variables in the 
formula.

\item \textbf{Canonical variable valuations.} Next, we demonstrate how 
to build a canonical structure of the variable valuations provided in Step 1. 
This canonical structure mirrors the canonical structure of \LTLFOC properties.

\item \textbf{Valuation of property instances.} A {\em property instance} is a 
unique substitution of variables in the property with values from their domains. 
This step demonstrate how to evaluate property instances.

\item \textbf{Applying quantifier numerical constraints.} This step 
demonstrates how to evaluate counting quantifiers by applying their numerical constraints 
on the valuation of a set of property instances from Step 3. The set of 
property instances is retrieved with respect to the canonical structure defined 
in Step 2.

\item \textbf{Inductive semantics.} Using the canonical structure in Step 2, 
and valuation of counting quantifiers in Step 4, we define semantics that begin at the 
outermost counting quantifier of an \LTLFOC property and evaluate quantifiers recursively 
inwards.
\end{enumerate}

\subsubsection{Variable Valuation}
We define a vector $\D$ with respect to a property $\varphi$ as follows: 
$$\D=\langle d_0,d_1,\cdots,d_{n-1}\rangle$$
where $n = |\BigQ|$ and $d_i$, $0 \leq i \leq n-1$, is a value for 
variable $x_i$. We denote the first $m$ components of the vector $\D$
(i.e., $\langle d_0,d_1,\cdots,d_{m-1}\rangle$) by $\D|^m$. We refer to $\D$ as 
a {\em value vector} and to $\D|^m$ as a {\em partial value vector}.


A {\em property instances} $\Fistar(\D|^m)$ is obtained by replacing 
every occurrence of the variables $x_0\cdots x_{m-1}$
in $\varphi$ with the values $d_0 \cdots d_{m-1}$, respectively. Thus, 
$\Fistar(\D|^m)$ is free of quantifiers of index
less than $m$, yet remains quantified over variables $x_m\cdots x_{n-1}$. For
instance, for the following property 
$$\varphi = \Forall_{>c_1}\,x: p_x(x) \Rightarrow
\left( \Forall_{<c_2} \,y : p_y(y) \Rightarrow \G\,q(x,y)\right)$$
and value vector $\D = \langle 1,2 \rangle$ (i.e., the vector of values for 
variables $x$ and $y$, respectively), 
$\Fistar(\D)$ will be $$\Fistar(\langle 1,2\rangle) =p_x(1) 
\Rightarrow \left(p_y(2)
\Rightarrow \G\,q(1,2)\right)$$

%

We now define the set $\BigD$ as the set of all value vectors with respect to a property $\varphi = \BigQ  \; \psi$ and a finite trace $u = 
u_0u_1\cdots u_k$:
\begin{equation}
	\label{eq:D} \BigD = \{\D \mid \exists j \in [0,k]:  
\forall i \in [0,n-1]: p_i(d_i) \in u_j\} 
\end{equation}
where $n=|\BigQ|$.

\subsubsection{Canonical Variable Valuations}
An \LTLFOC property follows a canonical structure, in which every counting quantifier
$\Q_i$ has a {\em parent} quantifier $\Q_{i-1}$, except for $\Q_0$ which is the
{\em root} counting quantifier. A counting quantifier $\Q_i$ is applied over all valuations of
its variable $x_i$ given a unique valuation of its predecessor variables
$x_0,\cdots ,x_{i-1}$.
Hence, we define function $\mathcal{P}$ which takes as 
input a partial value vector $\D|^m$, and returns all partial value vectors in 
$\BigD$ of length $m+1$, such that the first $m$ elements of these vectors is 
the same as $\D|^m$. In this context, we refer to $\D|^m$ as a {\em parent} 
vector and all the returned vectors as {\em child} vectors. Similarly, a property instance can have a parent; for instance, $\Fistar(\D|^m)$ is the parent of $\Fistar(\D|^{m+1})$.
\begin{equation*}
	\label{eq:eqP}
	\mathcal{P}(\varphi, u, \D|^m) = \bigg\{\D^\prime|^{m+1} \;\bigg|\; \D^\prime
\in \BigD \, \wedge \, \D^\prime|^m = \D|^m\bigg\}
\end{equation*}
Following the example above, assume there are two value vectors: $\langle 1,2\rangle$ and $\langle 1,3 \rangle$. In this case,
\begin{equation*}
	\label{eq:exP}
		\mathcal{P}(\varphi,u,\langle 1 \rangle) = \big\{\langle 1,2 \rangle,\langle 1,3\rangle\big\}
\end{equation*}

\subsubsection{Valuation of Property Instances}

As per the definition of $\BigD$, every value vector $\D=\langle 
d_0 \cdots d_{n-1}\rangle$ in $\BigD$ contains values for which the predicates 
$p_i(d_i)$ hold in some trace event $u_j$. For simplicity, we denote this as a 
value vector {\em in} a trace event $u_j$. These value vectors can possibly be 
in multiple and interleaved events in the trace. Thus, we define a trace 
$u^{\D}=u^{\D}_0 u^{\D}_1 \cdots u^{\D}_l$ as a subsequence of the trace $u$ 
such that the value vector $\D$ is in every event:
$$\forall \,j \in [0,l] : \forall \,i \in [0,n-1]: p_i(d_i) \in u^{\D}_j$$
For any property instance $\Fistar(\D)$, we wish to evaluate \linebreak $[u^{\D} \models_6 
\Fistar(\D)]$ (read as valuation of $\Fistar(\D)$ with respect to $u^{\D}$ 
for \LTLFOC), since 
any other event in trace $u$ is not of interest to $\Fistar(\D)$.

By leveraging $u^{\D}$, we define function $\mathcal{B}$ as follows: \\

\noindent$\mathcal{B}(\varphi,u,\D|^m,b) = $
\begin{equation*}
\begin{cases}
	\D^\prime|^{m+1} \in 
	\mathcal{P}(\varphi,u,\D|^m) \mid & \\
	\quad [u^{\D^\prime|^{m+1}}
\models_6 \Fistar(\D^\prime|^{m+1})] = b & \text{iff}~m<|\BigQ|-1 \\ 
	\D^\prime|^{m+1} \in 
	\mathcal{P}(\varphi,u,\D|^m) \mid & \\
	\quad [u^{\D^\prime|^{m+1}}
\models_4 \Fistar(\D^\prime|^{m+1})] = b & \text{iff}~m=|\BigQ|-1 \\ 
\end{cases}
\end{equation*} where $b$ is a truth value in $\mathbb{B}_6$. 
Function $\mathcal{B}$ can be implemented in a straightforward manner, where it 
iterates over all its children value vectors $\D^\prime|^{m+1}$ which are 
retrieved using $\mathcal{P}$. For every child vector, the function checks 
whether $\Fistar(\D^\prime|^{m+1})$ evaluates to $b$ with respect to the trace 
subsequence $u^{\D^\prime|^{m+1}}$.

To clarify $\mathcal{B}$, let us refer to our example earlier. Let a program trace $u$ be as follows:
$$u = \{p_x(1),p_y(2),\cdots\},\{p_x(1),p_y(3),\cdots\},\{p_x(1),p_y(2),\cdots\}$$
With respect to this trace, $\mathcal{P}(\varphi,u,\langle 1 \rangle) = \{\langle 
1,2\rangle,\langle 1,3 \rangle\}$. As per the definition of $u^{\D}$, 
$u^{\langle 1,2\rangle}=u_0u_2$, and $u^{\langle 1,3\rangle}=u_1$. Thus, 
$\mathcal{B}(\varphi,u,\langle 1 \rangle,b)$ checks the following:
\begin{align*}
	[u^{\langle 1,2\rangle}  &\models_4 p_x(1) \Rightarrow \left(p_y(2) \Rightarrow \G\,q(1,2)\right)] = b \\
	[u^{\langle 1,3\rangle}  &\models_4 p_x(1) \Rightarrow \left(p_y(3) \Rightarrow \G\,q(1,3)\right)] = b
\end{align*}
The definition of $u^{\D}$ implies that $p_i(d_i) \in u^{\D}_j$ for all $j$. 
Thus, we can simplify the property by omitting the $p$ predicates since they 
hold by definition:
\begin{align*}
	[u^{\langle 1,2\rangle}  &\models_4 \G\,q(1,2)] = b \\
	[u^{\langle 1,3\rangle}  &\models_4 \G\,q(1,3)] = b
\end{align*}
For instance, if only $[u^{\langle 1,2\rangle} \models_4 \G\,q(1,2)] = b$ holds, then 
$$\mathcal{B}(\varphi,u,\langle 1 \rangle,b) = \{\langle 1 ,2 \rangle\}$$

As can be seen in the example, the property instances that are evaluated are 
\LTLfour properties. This is because the input to $\mathcal{B}$ is 
$\D|^1=\D|^{|\BigQ|-1}$, which represents the inner most quantifier.

\subsubsection{Applying Quantifier Numerical Constraints}
Finally, numerical constraints should be incorporated in the semantics. 
We define function $\mathcal{S}$ as follows: 
\begin{equation}
\small
	\label{eq:S} 
	\mathcal{S}(\varphi,u,\D|^m,B) = 
	\begin{cases}
		\vphantom{\Bigg|}\bigg|\bigcup\limits_{b \in 
B}{\mathcal{B}(\varphi,u,\D|^m,b)}\bigg| \sim_i \\ \quad  c_i \times 
|\{\mathcal{P}(\varphi,u,\D|^m)\}| & \text{iff } Q_m = \Forall\\
		\vphantom{\Bigg|}\bigg|\bigcup\limits_{b \in B}{\mathcal{B}(\varphi,u,\D|^m,b)}\bigg| \sim_i c_i & \text{iff } Q_m = \Exists 
	\end{cases} 
\end{equation}
where $B \subseteq \mathbb{B}_6$ is a set of truth values. This function returns 
whether a counting quantifier constraint is satisfied or not based on any of the truth 
values $b \in B$. Observe that, for percentage counting quantifiers, the
constraint value denotes the percentage of property instances that evaluate to
$b$. For instance counting quantifiers, the constraint value denotes the number of
property instances that evaluate to $b$. For instance, consider Property~\ref{eq:example3} which is read as: for all users, there exists at most $3$ requests of type 
login that end with an unauthorized status. For such a property, if $4$ or more 
unauthorized login attempts are detected for the same user, the property is 
permanently violated.

%

\subsubsection{Inductive Semantics}
Using the previously defined set of of functions, we now formalize \LTLFOC 
semantics.
\begin{definition}[\LTLFOC Semantics]
	\label{def:semantics}
	 \LTLFOC semantics for properties with counting quantifiers are defined as 
follows: 
	\begin{equation*}
		\label{eq:semantics} [u \models_6 \varphi] = 
		\begin{cases}

\tru & \text{iff }\;\; \mathcal{S}(\varphi,u,\langle\rangle,\{\tru\}) = 1 \; 
\wedge \\ &\forall v \in \Sigma^{\omega} : [uv \models_6 \varphi] = \tru\\

\fals & 
\text{iff}\;\; \mathcal{S}(\varphi,u,\langle\rangle,\mathbb{B}_6-\{\fals\})= 0 
\; \wedge \\ &\forall v \in \Sigma^{\omega} : [uv \models_6 \varphi] = \fals\\

\TC & \text{iff}\;\; \mathcal{S}(\varphi,u,\langle\rangle,\{\tru,\TC\}) = 1 \; 
\wedge \\
			&\exists v \in \Sigma^{\omega} : [uv \models_6 \varphi] \neq \TC\\
			\FC & \text{iff }\;\; \mathcal{S}(\varphi,u,\langle\rangle,\mathbb{B}_6-\{\fals,\FC\}) = 0 \; \wedge \\
			&\exists v \in \Sigma^{\omega} : [uv \models_6 \varphi] \neq \FC \\
			\TT & \text{iff }\;\; \mathcal{S}(\varphi,u,\langle\rangle,\{\tru,\TC,\TT\}) = 1\; \wedge \\
			&\mathcal{S}(\varphi,u,\langle\rangle,\{\tru,\TC\}) = 0\\
			\FT & \text{iff }\;\; \mathcal{S}(\varphi,u,\langle\rangle,\{\tru,\TC,\TT\}) = 0\; \wedge \\
			
&\mathcal{S}(\varphi,u,\langle\rangle,\mathbb{B}_6-\{\fals,\FC\}) = 0 \qed
		\end{cases}
	\end{equation*}
\end{definition}

Note that these semantics are applied recursively until there is
only one counting quantifier left in the formula, at which point $\mathcal{B}$ checks
the valuation based on \LTLfour semantics ($[u^{\D} \models_4 \Fistar(\D)] = b$). When checking the valuation of these \LTLfour properties, $\mathcal{B}$ will always return an empty set 
in case the input $b$ is $\TC$ or $\FC$, since these truth values are 
inapplicable to \LTLfour properties. As mentioned earlier, truth values in 
$\mathbb{B}_6$ form a lattice. Standard lattice operators $\sqcap$ and $\sqcup$ 
are defined as expected based on the lattice's partial order.
		%
Permanent satisfaction $(\tru)$ or violation $(\fals)$ is applicable to 
$\Exists$ quantifiers regardless of the comparison operator, as well as a 
special case of $\Forall$ quantifiers:
\begin{itemize}
	\item \textbf{$\Forall$ quantifier.} As mentioned earlier, if the $\Forall$ quantifier is not subscripted, it is assumed to denote $\Forall_{=1}$. In this case, a single violation in its child property instances causes a permanent violation of the quantified property.
	\item \textbf{$\Exists$ quantifier.} Permanent violation is possible for any numerical constraint attached to an $\Exists$ quantifier, since it is a condition on the 
{\em number} of satisfied property instances. 
\end{itemize}

Property~\ref{eq:example3} illustrates an example of an $\Exists$ quantifier that can be permanently violated. Also, since the $\Forall$ quantifier in Property~\ref{eq:example3} defaults to $\Forall_{=1}$, it will be violated if a single user makes more than three unauthorized login attempts. In such a case, the entire property evaluates to $\fals$. Table~\ref{table:exists} illustrates how permanent satisfaction or violation apply to the different numerical constraints of $\Exists$ quantifiers.

\begin{equation}
	\label{eq:example3}
	\Forall x : \code{user}(x) \Rightarrow \left(\Exists_{\le3} \,r:\code{rid}(r) 
\Rightarrow \left( \code{login} \wedge \code{unauthorized} \right) \right)
\end{equation}

	\begin{table}
		[h!] \caption{Rules of permanent satisfaction or violation of $\Exists$ constraints} \label{table:exists} \centering 
		\begin{tabular}
			{c|l} Operator & Verdict \\
			\hline $> c$ & Permanent satisfaction if $> c$ \\
			$\ge c$ & Permanent satisfaction if $\ge c$ \\
			$= c$ & Permanent violation if $> c$ \\
			$< c$ & Permanent violation if $\ge c$ \\
			$\le c$ & Permanent violation if $> c$ \\
		\end{tabular}
	\end{table}
	
To clarify the semantics, consider Property~\ref{eq:example3} and the 
following program trace:
\begin{align*}
	&\{\code{rid}(12),\code{user}(Adam),\code{login},\code{unauthorized}\}\\
	&\{\code{rid}(13),\code{user}(Adam),\code{login},\code{unauthorized}\}\\
	&\{\code{rid}(14),\code{user}(Jack),\code{login},\code{authorized}\}\\
	&\{\code{rid}(15),\code{user}(Adam),\code{login},\code{unauthorized}\}\\
	&\{\code{rid}(16),\code{user}(Adam),\code{login},\code{unauthorized}\}
\end{align*}
where each line represents an event: a set of interpreted predicates. Each event contains a request identifier (\code{rid}), a username, a request type ($\code{login}$), and response status \linebreak ($\code{authorized}$ or $\code{unauthorized}$). As seen in the trace, there are $5$ distinct value vectors: $\langle Adam, 12\rangle$, $\langle Adam, 13\rangle$, $\langle Jack, 14\rangle$, \linebreak $\langle Adam, 15\rangle$, and $\langle Adam, 16\rangle$. Now, let us apply the inductive semantics on the property.

\textbf{Step 1.} We begin by checking the truth value of $[u \models_6 \varphi]$, 
which requires determining which condition in Definition~\ref{def:semantics} 
applies. This requires the evaluation of function $\mathcal{S}$ for the 
different truth values shown. Since we are verifying $\varphi$, we begin with 
the outermost counting quantifier, which is a $\Forall$ quantifier. Thus, $\mathcal{S}$ 
will require calculating the cardinality of the set 
$\mathcal{P}(\varphi,u,\D|^0)$, which in case of the trace should be 
$|\{Adam,Jack\}| = 2$. Now, in order to evaluate $\mathcal{S}$, one has to 
evaluate $\mathcal{B}$ to determine whether each property instance 
evaluates to a certain truth value or not. The two property instances thus far 
are:
	\begin{align*}
		\Fistar(\D|^1)&=\Fistar(Adam)=\Exists_{\le3} \,r:\code{rid}(r) \Rightarrow \left( 
 \code{login} \wedge \code{unauthorized} \right) \\
		\Fistar(\D^\prime|^1)&=\Fistar(Jack)=\Exists_{\le3} \,r:\code{rid}(r) \Rightarrow \left( 
 \code{login} \wedge \code{unauthorized} \right)
	\end{align*} 
And the trace subsequences for these property instances respectively are:
	\begin{align*}
		u^{\D|^1}=&\{\code{rid}(12),\cdots\}\{\code{rid}(13),\cdots\}\{\code{rid}(15),\cdots\}\{\code{rid}(16),\cdots\} \\
		u^{\D^\prime|^1}=&\{\code{rid}(14),\cdots\}
	\end{align*} 
Note that $\code{user}(Adam) \Rightarrow \cdots$ is omitted from 
$\Fistar(\D|^1)$ since $\code{user}(Adam)$ holds according to the trace 
subsequence. The same applies to $\code{user}(Jack)$. Evaluating these property 
instances with respect to the trace subsequences requires referring to 
Definition~\ref{def:semantics} again, which marks the second level of recursion.

\textbf{Step 2.} Let us consider the property instance $\Fistar(\D|^1)$, which
begins with an $\Exists$ quantifier and has \LTLfour properties as child
instances (refer to $\mathcal{P}$). These properties are in the form of
$\code{login}\, \wedge\, \code{unauthorized}$, where there is one
instance for each distinct request identifier. We can deduce that the property
holds for all $4$ requests: $12$, $13$, $15$, and $16$, thus evaluating to
$\tru$. Therefore, the following holds:
$$\mathcal{B}(\Fistar(\D|^1),u^{\D|^1},\D|^1,\tru) = 4$$
This value, when used in $\mathcal{S}(\Fistar(\D|^1),u^{\D|^1},\D|^1,\{\tru\})$ will violate the numerical condition: $4 \not\le 3$, resulting in $\mathcal{S}$ valuating to $0$ (false). Based on the conditions in Definition~\ref{def:semantics} and the rules of permanent violation, this property instance becomes permanently violated and thus the verdict is $\fals$.

The other property instance $\Fistar(\D^\prime|^1)$ will however evaluate to  $\tru$ since its child property instance $$\Fistar(\D^\prime|^2)=\Fistar(\langle Jack,14\rangle) =
 \code{login} \,\wedge\, \code{unauthorized}$$ is violated, and thus the number of satisfied instances is still less than $3$.

\textbf{Step 3.} In this step we use the valuations determined in Step 2 to produce verdicts for the property instances in Step 1. Based on $\mathcal{S}$, the $\Forall$ quantifier's numerical condition is violated, since not {\em all} instances are satisfied. The final verdict should thus be $[u \models_6 \varphi] = \fals$, which denotes a permanent violation of the property.
%

%% file: monitor.tex
\vspace{-3mm}
\section{Divide-and-Conquer-based \\ Monitoring of LTL4-C}
\label{sec:monitor}

In this section, we describe our technique inspired by divide-and-conquer for 
evaluating \LTLFOC properties at run time. This approach forms the basis of our 
parallel verification algorithm in Section~\ref{sec:design}.

Unlike runtime verification of propositional \LTLfour properties, where the 
structure of a monitor is determined solely based on the property itself, a 
monitor for an \LTLFOC needs to evolve at run time, since the valuation of 
quantified variables change over time. More specifically, the monitor 
$\mathcal{M}_\varphi$ for an \LTLFOC property $\varphi = \BigQ \psi$ relies on 
instantiating a {\em submonitor} for each property instance $\Fistar$ obtained 
at run time. We incorporate two type of submonitors: (1) {\em \LTLfour 
submonitors} evaluate the inner \LTL property $\psi$, and (2) {\em quantifier 
submonitors} deal with quantifiers in $\BigQ$, described in 
Subsections~\ref{subsec:ltl4sub} and~\ref{subsec:quantsub}. In 
Subsection~\ref{subsec:instant}, we explain the conditions under which a 
submonitor is instantiated at run time. Finally, in 
Subsection~\ref{subsec:truthsub}, we elaborate on how submonitors evaluate an 
\LTLFOC property.

\subsection{LTL4 Submonitors}
\label{subsec:ltl4sub}

Let $\varphi = \BigQ \psi$ be an \LTLFOC property. If $|\BigQ| = 0$ 
(respectively, one wants to evaluate $\Fistar(\D|^i)$, 
where $i=|\BigQ|$), then $\varphi$ (respectively, $\Fistar(\D|^i)$) is free of 
quantifiers and, thus, the monitor (respectively, submonitor) of such a 
property is a standard \LTLfour monitor (see Definition~\ref{def:monitor2}). We 
denote \LTLfour submonitors as $\mathcal{M}^*_{\D}$, where $\D$ is the value 
vector with which the monitor is initialized.

\subsection{Quantifier Submonitors}
\label{subsec:quantsub}

Given a finite trace $u$ and an \LTLFOC property $\varphi = \BigQ \psi$, a {\em 
quantifier submonitor} ($\mathcal{M}^{\Q}$) is a monitor responsible for determining the valuation of a 
property instance $\Fistar(\D|^i)$ with respect to a trace subsequence 
$u^{\D|^i}$, if $i < |\BigQ|$. 
Obviously, such a valuation is in $\mathbb{B}_6$. Let $\mathbb{V}$ be a 
six-dimensional vector space, where each dimension represents a truth value in 
$\mathbb{B}_6$.
\begin{definition} [Quantifier Submonitor]
\label{def:qmonitor}
Let \linebreak $\varphi=\BigQ\psi$ be an \LTLFOC property and $\Fistar(\D|^i)$ be a 
property instance, with $i \in [0,|\BigQ|-1]$. The {\em quantifier submonitor} for 
$\Fistar(\D|^i)$ is the tuple $\mathcal{M}_{\D|^i}^{\Q} = \langle 
\Q_i,\mathbb{M}_{\D|^i},v,b\rangle$, where
\begin{itemize}
	\itemsep0em
 \item $\Q_i$ encapsulates the quantifier information (see 
Equation~\ref{eq:bigq})
\item $v \in \mathbb{V}$ represents the current number of child property 
instances that evaluate to each truth value in $\mathbb{B}_6$ with respect to 
their trace subsequences,
\item $b \in \mathbb{B}_6$ is the current value of ${[u^{\D|^i} 
\models_6 \Fistar(\D|^i)]}$,
\item $\mathbb{M}_{\D|^i}$ is the set of child 
submonitors (submonitors of child property instances) defined as follows:
\begin{equation*}
	\mathbb{M}_{\D|^i} = \begin{cases}
		\{\mathcal{M}^*_{\D^\prime} \mid \D^\prime|^i=\D|^i\} & 
\text{if } i = |\BigQ|-1 \\
		\{\mathcal{M}^{\Q}_{\D^\prime|^{i+1}} \mid \D^\prime|^i=\D|^i\} 
& \text{if } i < |\BigQ| - 1
	\end{cases}
\end{equation*}
\end{itemize}
  Thus, if $i = |\BigQ| - 1$, all child submonitors are \LTLfour submonitors. 
Otherwise, they are quantifier submonitors of the respective child property 
instances.\qed
\end{definition}

Based on the definition, every quantifier submonitor references a set of child monitors. We use the following notation to denote a hierarchy of a submonitor and its children:
$$\mathcal{M}^{\Q}_{\D|^i} \big\{ \mathcal{M}^{\Q}_{\D|^{i+1}}, \mathcal{M}^{\Q}_{\D^\prime|^{i+1}}, \mathcal{M}^{\Q}_{\D^{\prime\prime}|^{i+1}},\cdots\big\}$$ such that $\D|^i=\D^\prime|^i = \D^{\prime\prime}|^{i}\cdots$ and $i < |\BigQ|-1$ which is why the child monitors are quantifier submonitors.
%
%

\subsection{Instantiating Submonitors}
\label{subsec:instant}

Let an \LTLFOC monitor $\mathcal{M}_\varphi$ for property $\varphi$ evaluate the property with respect to a finite trace $u=u_0u_1 \cdots$. Let $\D=\langle d_0,d_1,\cdots\rangle$ be a value vector and $u_k$ the first trace event such that $\forall d_i : p_i(d_i) \in u_k$, where $p_i$ is the predicate within each quantifier (i.e. $\Forall x_i : p_i(x_i) \Rightarrow \cdots$). In this case, the \LTLFOC 
monitor instantiates submonitors for every property instance resulting from that 
value vector. A value vector of length $|\BigQ|$ results in $|\BigQ|+1$ property 
instances: one for each quantifier in addition to an \LTLfour inner property.  
The hierarchy of the instantiated submonitors is as follows:
$$\mathcal{M}^{\Q}_{\D|^0} \bigg\{\mathcal{M}^{\Q}_{\D|^1} \Big\{ \cdots \big\{ \mathcal{M}^{\Q}_{\D|^{|\BigQ-1|}} \{\mathcal{M}^{*}_{\D}\}\big\}\Big\}\bigg\}$$

If another value vector $\D^\prime$ is subsequently encountered for the first time, the hierarchy of submonitors becomes as follows:
$$\mathcal{M}^{\Q}_{\D|^0} \Big\{\mathcal{M}^{\Q}_{\D|^1} \big\{ \cdots  \{\mathcal{M}^{*}_{\D}\}\big\}, \mathcal{M}^{\Q}_{\D^\prime|^1} \big\{ \cdots  \{\mathcal{M}^{*}_{\D^\prime}\}\Big\}$$
Since the hierarchy is formulated as a recursive set, no 
duplicate submonitors are allowed. Two submonitors are duplicates, if they 
represent identical value vectors. If $\D|^1 = \D^\prime|^1$, 
the respective monitors are merged. Such merging is explained in detail in 
Section~\ref{sec:design}.

\subsection{Evaluating LTL4-C Properties}
\label{subsec:truthsub}
Once the \LTLFOC monitor instantiates its submonitors, every submonitor is 
responsible for updating its truth value. The truth value of \LTLfour submonitors
($\mathcal{M}^*$) maps to the current state of the submonitor's automaton as
described in Definition~\ref{def:monitor2}. Quantifier submonitors update their
truth value based on the truth values of all child submonitors. The number of
child submonitors whose truth value is $\tru$ is stored in $v_{\tru}$ (i.e., the 
$\tru$ dimension of vector $v$) and so on
for all truth values in $\mathbb{B}_6$. Then, \LTLFOC semantics are applied,
beginning with function $\mathcal{S}$ (see Equation~\ref{eq:S}), which in turn
relies on the cardinality of function $\mathcal{B}(\varphi,u,\D|^{i},b)$ where
$b$ is a truth value. This cardinality is readily provided by the vector $v$,
such that for instance $\mathcal{B}(\varphi,u,\D|^{i},\tru) = v_\tru$ and so on.

\begin{figure}[t]
	\centering 
	\includegraphics[scale=.5]{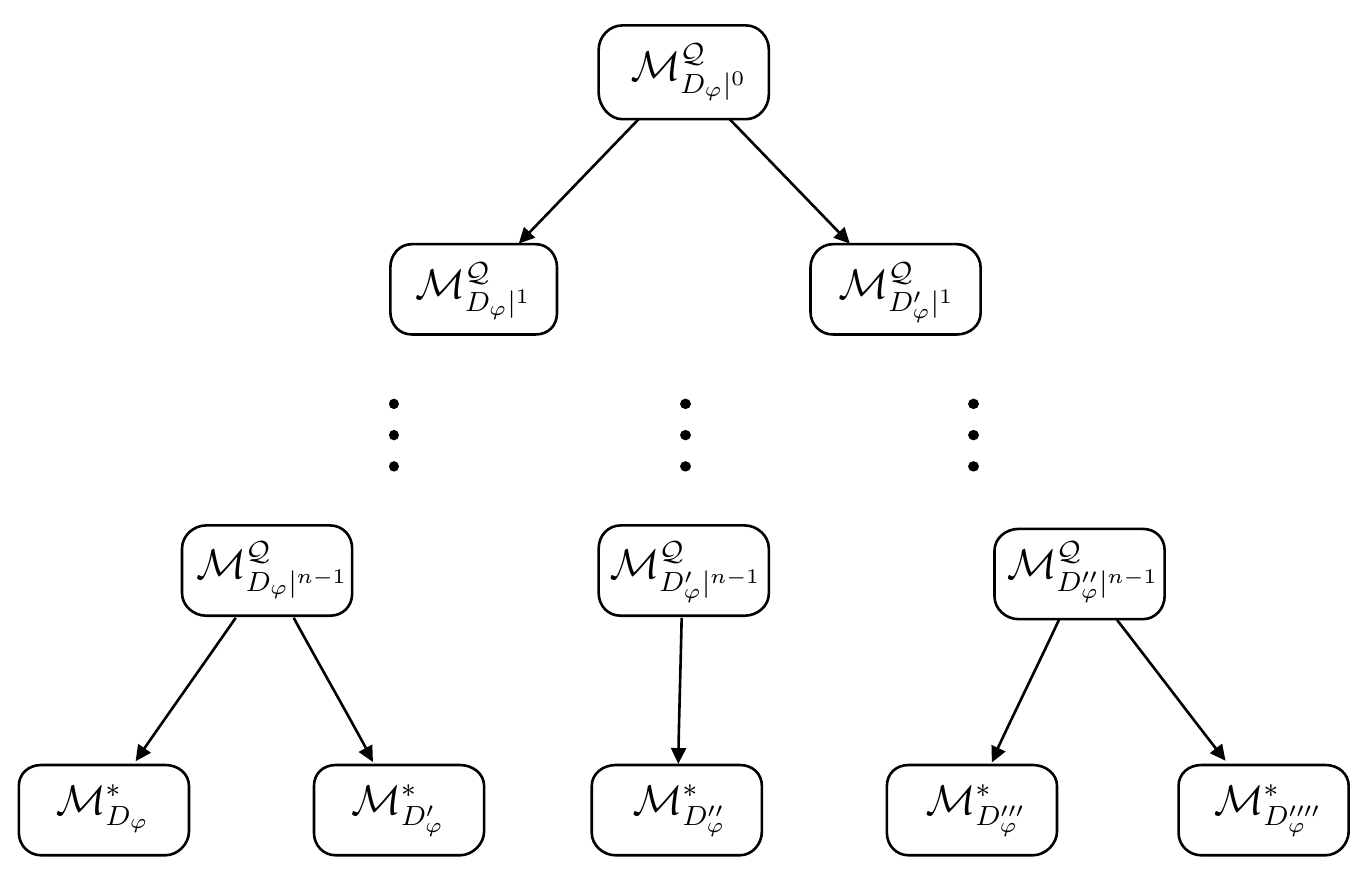} 
	\caption{Tree structure of an \LTLFOC monitor.} 
	\label{fig:tree}
	\vspace{-3mm}
\end{figure}

Since each submonitor depends on its child submonitors, updating truth values proceeds
outwards, starting at \LTLfour submonitors, then recursively parent submonitors update their truth values
until the submonitor $\mathcal{M}^{\Q}_{\D|^0}$, which is the root submonitor. 
The truth value of the root submonitor is the truth value of property $\varphi$ 
with respect to trace $u$. This is visualized as the tree shown in 
Figure~\ref{fig:tree}.

%% file: design.tex

\vspace{-3mm}
\section{Parallel Algorithm Design}
\label{sec:design}

The main challenge in designing a runtime monitor is to ensure that its 
behavior does not intervene with functional and extra-functional (e.g., 
timing constraints) behavior of the program under scrutiny. This section 
presents a parallel algorithm for verification of \LTLFOC properties. Our 
idea is that such a parallel algorithm enables us to offload the monitoring 
tasks into a different computing unit (e.g., the GPU). The algorithm utilizes 
the popular {\em MapReduce} technique to spawn and merge submonitors to 
determine the final verdict. This section is organized as follows: Subsection~\ref{subsec:valext} describes how valuations are extracted from a trace in run time, and Subsection~\ref{subsec:algsteps} describes the steps of the algorithm in detail.

\subsection{Valuation Extraction}
\label{subsec:valext}
Valuation extraction refers to obtaining a valuation of quantified variables 
from the trace. As described in \LTLFOC semantics, the predicate $p_i(x_i)$ 
identifies the subset of the domain of $x_i$ over which the quantifier is 
applied: namely the subset that exists in the trace. From a theoretical 
perspective, we check whether the predicate is a member of some trace event, 
which is a set of predicates. From an implementation perspective, the trace 
event is a key-value structure, where the key is for instance a string 
identifying the quantified variable, and the value is the concrete value of the 
quantified variable in that trace event. Consider the following property: 
\begin{equation}
\label{eq:prop}
\varphi=\Forall_{\ge 0.95} \, s \, \text{:}\, \code{socket}(s) \Rightarrow \left( \G \,\code{recv}\left(s\right) \Rightarrow \textbf{F} \, \code{respond}\left(s\right) \right)
\end{equation}
Predicate $p$ in this case is $\code{socket}(s)$, and a trace event should 
contain a key $\code{socket}$ and a value $\in [0,65535]$ representing the 
socket file descriptor in the system. Thus, the valuation extraction function 
$\varepsilon(u_i,K)=\D$ returns a map where keys are in $K$, and the value of 
each key is the value of the quantified variable corresponding to this key. 
These keys are defined by the user.

\subsection{Algorithm Steps}
\label{subsec:algsteps}
Algorithm~\ref{alg:monitor} presents the pseudocode of the parallel monitoring 
algorithm. Given an \LTLFOC property $\varphi=\BigQ\, \psi$, the input to the 
algorithm is the \LTLfour monitor $\mathcal{M}^*$ of \LTLfour property $\psi$, a 
finite trace $u$, the set of quantifiers $\BigQ$, and the vector of keys $K$ 
used to extract valuations. Note that the algorithm supports both online and offline runtime verification. Offline mode is straightforward since the algorithm receives a finite trace that it can evaluate. In the case of online mode, the algorithm maintains data structures that represent the tree structure shown in Figure~\ref{fig:tree}, and repeated invocation of the algorithm updates these data structures incrementally. Thus, a monitoring solution can invoke the algorithm periodically or based on same event in an online fashion, and still receive an evolving verdict.

The entry point to the algorithm is at 
Line~\ref{line:sort} which is invoked when the monitor receives a trace to 
process. The algorithm returns a truth value of the property at 
Line~\ref{line:applyq}. Subsections~\ref{subsubsec:sort} -- 
\ref{subsubsec:apply} describe the functional calls between 
Lines~\ref{line:sort} -- \ref{line:applyq}. The MapReduce operations are
visible in functions {\em SortTrace} and {\em ApplyQuantifiers}, which perform
a {\em map} ($\rightrightarrows$) in Lines~\ref{line:getval} and~\ref{line:getc} respectively. {\em
ApplyQuantifiers} also performs a reduction ($\rightarrowtail$) in Line~\ref{line:reduce}.

\subsubsection{Trace Sorting}
\label{subsubsec:sort}
As shown in Algorithm~\ref{alg:monitor}, the first step in the algorithm is to 
sort the input trace $u$ (Line~\ref{line:sort}). The function {\em SortTrace} performs this 
functionality as follows:
\begin{enumerate}
	\item The function performs a parallel map of every trace event to the value vector that it holds using $\varepsilon$ (Line~\ref{line:getval}).
	\item The mapped trace is sorted in parallel using the quantifier variable keys (Line~\ref{line:sortk}). For instance, according to Property~\ref{eq:prop}, the key used for sorting will be $\code{socket}$, effectively sorting the trace by socket identifier.
	\item The sorted trace is then compacted based on valuations, and the function returns a map $\mu$ where keys are value vectors and values are the ranges of where these value vectors exist in trace $u$ (Line~\ref{line:compact}). A range contains the start and end index. This essentially defines the subsequences $u^{\D}$ for each property instance $\Fistar(\D)$ (refer to Subsection~\ref{subsec:semantics}).
\end{enumerate}

\subsubsection{Monitor Spawning}
\label{subsubsec:spawning}
Monitor spawning is the second step of the algorithm (Line~\ref{line:spawn}). The function
{\em SpawnMonitors} receives a map $\mu$ and searches the cached collection of
previously encountered value vectors $\mathbb{D}$ for duplicates. If a value
vector in $\mu$ is new, it creates submonitors and inserts them in the tree of
submonitors $T$ (Line~\ref{line:addt}). The function {\em AddToTree} attempts to generate
$|\BigQ|-1$ quantifier submonitors $\mathcal{M}^{\Q}$ (Line~\ref{line:submon}) ensuring there
are no duplicate monitors in the tree (Line~\ref{line:nodup}). After all quantifier
submonitors are created, {\em SpawnMonitors} creates an \LTLfour submonitor
$\mathcal{M}^{*}$ and adds it as a child to the leaf quantifier submonitor in
the tree representing the value vector (Line~\ref{line:addm}). This resembles the
structure in Figure~\ref{fig:tree}. Creation of submonitors is performed in
parallel for all value vectors in trace $u$.

\subsubsection{Distributing the Trace}
The next step in the algorithm is to distribute the sorted trace to all \LTLfour submonitors (Line~\ref{line:dist}). The function {\em Distribute} instructs every \LTLfour submonitor to process its respective trace by passing the full trace and the range of its respective subsequence, which is provided by the map $\mu$ (Line~\ref{line:procbuf}). The \LTLfour monitor updates its state according to the trace subsequence and stores its truth value $b$.

\subsubsection{Applying Quantifiers}
\label{subsubsec:apply}
Applying quantifiers is a recursive process, beginning with the leaf quantifier 
submonitors and proceeding upwards towards the root of the tree 
(Line~\ref{line:applyq}). Function {\em ApplyQuantifiers} operates in the 
following steps:
\begin{enumerate}
	\item The function retrieves all quantifier submonitors at the $i^{th}$ level in the tree $T$ (Line~\ref{line:treelevel}).
	\item In parallel, for each quantifier submonitor, all child submonitor truth values are reduced into a single truth value of that quantifier submonitor (Lines~\ref{line:getc}-\ref{line:getb}). This step essentially {\em reduces} all child truth vectors into a single vector and then applies \LTLFOC semantics to determine the truth value of the current submonitor.
	\item The function proceeds recursively calling itself on submonitors that are one level higher. It terminates when the root of the tree is reached, where the truth value is the final verdict of the property with respect to the trace. 
\end{enumerate}

\input{Algorithm1.tex}

%% file: Algorithm1.tex
\algrenewcommand\alglinenumber[1]{\scriptsize #1:}
\begin{algorithm}
\scriptsize
\begin{algorithmic}[1]
\State INPUT: An \LTLfour monitor $\mathcal{M}^*$ of \LTL property $\psi$, a 
finite trace $u$, a set of 
quantifiers $\BigQ$, and a vector of keys $K$ to extract valuations of 
quantified variables.\vspace{1mm}
\State declare $T=\{\mathcal{M}^{\Q}_{D|^0}\}$ \Comment{Tree of quantifier submonitors}
\State declare $\mathbb{D}=\{\}$ \Comment{Value vector set}
\State declare $\mathbb{M}^*=\{\}$ \Comment{\LTLfour submonitor 
set}
\State $\mu \gets$ \Call{SortTrace}{$u$} \Comment{The entry point} \label{line:sort}
\State \Call{SpawnMonitors}{$\mu$} \label{line:spawn}
\State \Call{Distribute}{$u$,$\mu$} \label{line:dist}
\State return \Call{ApplyQuantifiers}{$|\BigQ-1|$} \label{line:applyq}
\par\noindent\rule{8cm}{1pt}
\Function{SortTrace}{$u$}\Comment{Trace sorting and compaction}
	\State $u_i \rightrightarrows u^\prime_i := \varepsilon(u_i,K)$  \Comment{$\parallel$ map to value vectors} \label{line:getval}
	\State \Call{ParallelSort}{$u^\prime$,$K$}\label{line:sortk}
	\State $\mu\langle D,r\rangle \gets $ \Call{ParallelCompact}{$u^\prime$}\label{line:compact}
	\State return $\mu$
\EndFunction
\par\noindent\rule{8cm}{1pt}
\Function{SpawnMonitors}{$\mu$}\Comment{Monitor spawning}
\For{$D \in \mu$} \textbf{in parallel}
	\If{$D \not\in \mathbb{D}$}
		\State \Call{Add}{$\mathbb{D}$,$D$}
		\State $t \gets$ \Call{AddToTree}{$D$} \label{line:addt}
		\State $t$.addMonitor(\Call{CreateMonitor}{$D$})\label{line:addm}
	\EndIf
	\EndFor
\EndFunction
\par\noindent\rule{8cm}{1pt}
\Function{AddToTree}{$D$}
\State $t$ = $T$.root
\For{$i \in [1,|\BigQ|-1]$} \label{line:submon}
	\If{$\mathcal{M}^{\Q}_{D|^i} \not\in t$.children} \label{line:nodup}
		\State $t$.addchild($\mathcal{M}^{\Q}_{D|^i}$)
	\EndIf
	\State $t \gets t$.children$\left[\mathcal{M}^{\Q}_{D|^i}\right]$
\EndFor
\State return $t$
\EndFunction
\par\noindent\rule{8cm}{1pt}
\Function{CreateMonitor}{$D$}\Comment{Monitor creation}
\State $\mathcal{M}^*_{D} \gets $ \Call{LaunchMonitorThread}{$D$}
\State $\mathcal{M}^*_{D}.D \gets D$
\State \Call{add}{$\mathbb{M}^*$,$\mathcal{M}^*_{D}$}
\State return $\mathcal{M}^*_{D}$
\EndFunction
\par\noindent\rule{8cm}{1pt}
\Function{Distribute}{$u$,$\mu$}\Comment{Distribute trace to monitors}
\For{$\mathcal{M}^*_{D} \in \mathbb{M}^*$} \textbf{in parallel}
	\State \Call{ProcessBuffer}{$\mathcal{M}^*_{D}$,$u$,$\mu[\mathcal{M}^*_{D}.D]$} \label{line:procbuf}
	\EndFor
\EndFunction
\par\noindent\rule{8cm}{1pt}
\Function{ProcessBuffer}{$\mathcal{M}^*_{D}$,$u$,$r$}\Comment{Process trace subsequence}
\State filter include $u \rightrightarrows u^\prime := u[r.\text{start}, r.\text{end}]$ \Comment{$\parallel$ filter}
\State $\mathcal{M}^*_{D}.b$ $ \gets $\Call{UpdateMonitor}{$\mathcal{M}^*_{D}$, $u^\prime$}
\EndFunction
\par\noindent\rule{8cm}{1pt}
\Function{ApplyQuantifiers}{$i$}\Comment{Apply quantifiers}
\For{$t \in T.$nodesAtDepth($i$)} \textbf{in parallel} \label{line:treelevel}
	\State $t.$children $\rightrightarrows  \{s:=[v,v^\prime,\cdots]\}$\Comment{$\parallel$ map to truth vectors} \label{line:getc}
	\State $s \rightarrowtail t.v$ \Comment{$\parallel$ reduction to truth vector} \label{line:reduce}
	\State $t.b \gets $ \Call{Valuation}{$t$} \Comment{\LTLFOC semantics} \label{line:getb}
\EndFor
\If{$i=0$}
	\State return $t.b$
\EndIf
\State return \Call{ApplyQuantifiers}{$i-1$}
\EndFunction
%
%
%
%

\end{algorithmic}
\normalsize
\caption{\LTLFOC monitoring algorithm}
\label{alg:monitor}
\end{algorithm}

%% file: experiments.tex
\vspace{-3mm}
\section{Implementation and \\ Experimental Results}
\label{sec:exp}


We have implemented Algorithm~\ref{alg:monitor} for two computing technologies: 
Multi-core CPUs and GPUs. We applied three optimizations in our GPU-based 
implementation: (1) we use {\em CUDA Thrust API} to implement parallel 
sort, (2) we using {\em Zero-Copy Memory} which parallelizes
data transfer with kernel operation without caching, and (3) we 
enforced alignment, which enables coalesced read of trace events into monitor
instances. In order to intercept systems calls, we have integrated our algorithm 
with the Linux \texttt{strace} application, which logs all system calls made by 
a process, including the parameters passed, the return value, the time the 
call was made, etc. Notice that using \texttt{strace} has the 
benefit of eliminating static analysis for 
instrumentation. The work in~\cite{caceres2002syscall, wang2004file, 
ramsbrock2007profiling} also use {\tt strace} to debug the behavior of 
applications.

Subsection~\ref{subsec:casestudies} presents the case studies implemented to 
study the effectiveness of the GPU implementation in online and offline
monitoring. Subsection~\ref{subsec:setup} discusses the experimental setup, 
while Subsection~\ref{subsec:results} analyzes the results.

\subsection{Case studies}
\label{subsec:casestudies}
We have conducted the following three case studies:
\vspace{-2mm}
\begin{enumerate}
	\itemsep0em
	\item \textbf{Ensuring every request on a socket is responded to.} This 
case study monitors the responsiveness of a web server. Web servers under heavy 
load may experience some timeouts, which results in requests that are not 
responded to. This is a factor contributing to the uptime of the server, along 
with other factors like power failure, or system failure. Thus, we monitor that 
at least $95\%$ of requests are indeed responded:
\begin{equation*}
\Forall_{\ge 0.95} \, s \, \text{: } \code{socket}(s)\G\, 
\code{receive}\left(s\right) \Rightarrow \textbf{F} \, 
\code{respond}\left(s\right)
\end{equation*}
We utilize the Apache Benchmarking tool to generate different load 
levels on the Apache Web Server.

\item \textbf{Ensuring fairness in utilization of personal cloud storage 
services.} This case study is based on the work in~\cite{drago2012inside}, which 
discusses how profiling DropBox traffic can identify the bottlenecks and 
improve the performance. Among the issues detected during this analysis, 
is a user repeatedly uploading chunks of maximum size to DropBox servers. Thus, it is beneficial for a runtime verification system to ensure that the average chunk size of all clients falls below a predefined maximum threshold, effectively ensuring fairness of service use. The corresponding \LTLFOC property is as follows:
	\begin{equation*}
		\Forall u \, \text{: } \code{user}(u) \Rightarrow \textbf{F} 
\, (\code{avg\_chunksize}\left(u\right) \le \code{maximum})
	\end{equation*}
where $\code{avg\_chunksize}$ is a predicate that is based on a variable in the program representing the average chunk size of the current user's session.
\item \textbf{Ensuring proxy cache is functioning correctly.} This 
experiment is based on a study that shows the effectiveness of utilizing proxy 
cache in decreasing \linebreak YouTube videos requests in a large university 
campus~\cite{zink2008watch}. Thus, we monitor that no video is requested 
externally while existing in the cache:
		\begin{equation*}
			\Forall v : \code{vid}(v) \Rightarrow \Exists_{=0}\, r : 
\code{req}(r) \Rightarrow (\code{cached}(v)\, \wedge\, \code{external}(r))
		\end{equation*}
\end{enumerate}

\subsection{Experimental Setup}
\label{subsec:setup}
\noindent \textbf{Experiment Hardware and Software.} The machine we use to run 
experiments comprises of a 12-core Intel Xeon E5-1650 CPU, an 
Nvidia Tesla K20c GPU, and $32$GB of RAM, running Ubuntu 12.04.\\

\noindent \textbf{Experimental Factors.} The experiments involve comparing the 
following factors:
\begin{itemize}
	\setlength{\itemsep}{-1pt} 
	\item {\em Implementation.} We compare three implementations of the \LTLFOC monitoring algorithm:
\vspace{-2mm}
	\begin{itemize}
	\itemsep0em
		\item {\em Single Core CPU.} A CPU implementation running on a single core. The justification for using a single core is to allow the remaining cores to perform the main functionality of the system without causing contention from the monitoring process.
		\item {\em Parallel CPU.} A CPU implementation running on all 12 cores of the system. The implementation uses OpenMP.
		\item {\em GPU.} A parallel GPU-based implementation.
	\end{itemize}
	\item {\em Trace size.} We also experiment with different trace sizes to study the scalability of the monitoring solution, increasing exponentially from $16,384$ to $8,388,608$.
\end{itemize}

\noindent \textbf{Experimental Metrics.} Each experiment results in values for the following metrics:
\vspace{-2mm}
\begin{itemize}
	\itemsep0em
	\item {\em Total execution time.} The total execution time of the 
monitor. 
	\item {\em Monitor CPU utilization.} The CPU utilization of the monitor process.
\end{itemize}
In addition, we measure the following metrics for Case Study 1, 
since it utilizes an online monitor:
\vspace{-2mm}
\begin{itemize}
	\setlength{\itemsep}{-2pt} 
	\item {\em Monitored program CPU utilization.} The CPU utilization of the monitored program. This is to demonstrate the impact of monitoring on overall CPU utilization.
	\item {\tt strace} {\em parsing CPU utilization.} The CPU 
utilization of the \code{strace} parsing module. This module translates 
\code{strace} strings a numerical table.
\end{itemize}
We perform $20$ replicates of each experiment and present error bars of a $95\%$ confidence interval.
\subsection{Results}
\label{subsec:results}

%

\begin{figure}[t]
	\centering 
	\includegraphics[scale=.45]{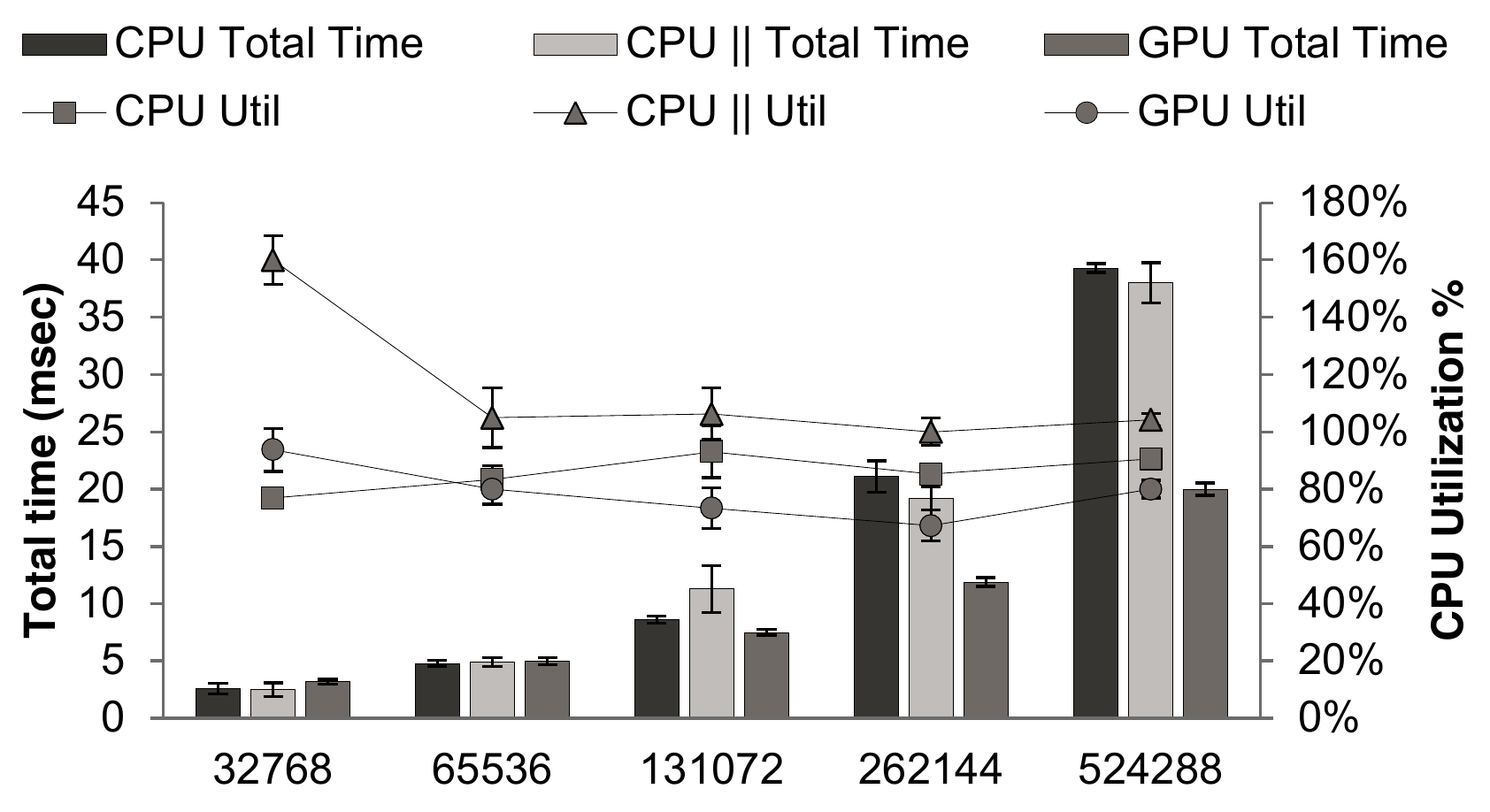} 
	\caption{Results of Case Study 1.} 
	\label{fig:strace} 
	\vspace{-1mm}
\end{figure}

\begin{figure}[t]
	\centering 
	\includegraphics[scale=.45]{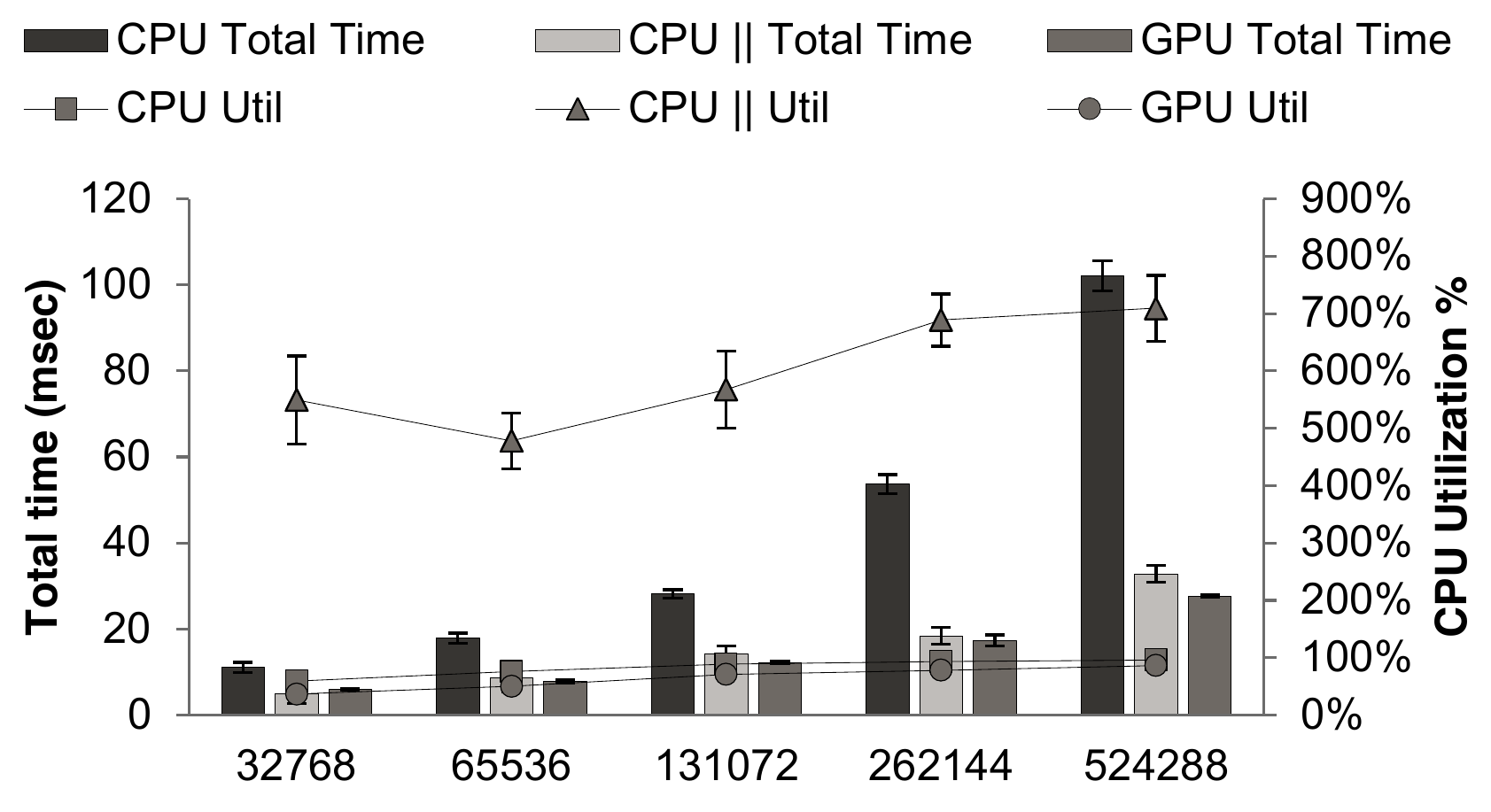} 
	\caption{Results of Case Study 2.} 
	\label{fig:dropbox} 
	\vspace{-5mm}
\end{figure}

\begin{figure}[t]
	\centering 
	\includegraphics[scale=.45]{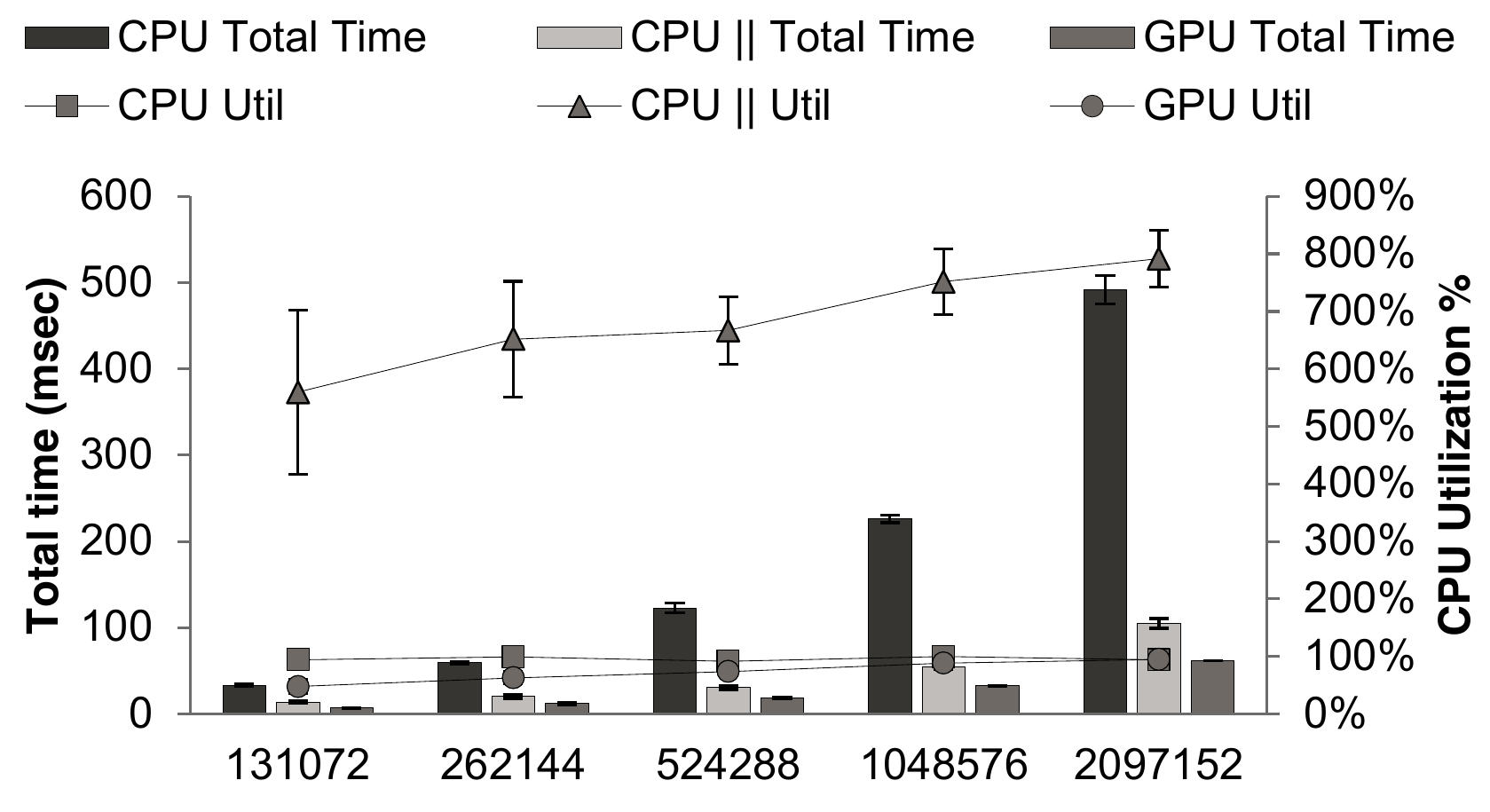} 
	\caption{Results of Case Study 3.} 
	\label{fig:youtube} 
	\vspace{-5mm}
\end{figure}

	The results of Case Study 1 are shown in Figure~\ref{fig:strace}. As seen
in the figure, the GPU implementation scales efficiently with increasing trace
size, resulting in the lowest monitoring time of all three implementations. The
GPU versus single core CPU speedup ranges from $0.8$ to $1.6$, increasing with
the increasing trace size. When compared to parallel CPU (CPU ||), the speedup
ranges from $0.78$ to $1.59$. This indicates that parallel CPU outperforms GPU
for smaller traces ($32768$), yet does not scale as well as GPU in this case study. This is attributed to the low number of individual objects in the trace, making parallelism less impactful. CPU
utilization results in Figure~\ref{fig:strace} show a common trend with the
increase of trace size. When the trace size is small, parallel implementations
incur high CPU utilization as opposed to a single core implementation, which could be
attributed to the overhead of parallelization relative to the small trace size.
On the other hand, GPU shows a stable utilization percentage, with a $78\%$
average utilization. The single core CPU implementation shows a similar trend,
yet slightly elevated average utilization (average $86\%$). The parallel CPU
implementation imposes a higher CPU utilization (average $1.15\%$), since more
cores are being used to process the trace. This result indicates that shipping
the monitoring workload to GPU consistently provides more time for CPU to
execute other processes including the monitored process. The results of Case
Study 2 and Case Study 3 in Figures~\ref{fig:dropbox} and~\ref{fig:youtube}
respectively provide a different perspective. The number of individual objects in these traces are large, making parallelism highly effective. For Case Study 2, the speedup of the GPU
implementation over single core CPU ranges from $1.8$ to $3.6$, and $0.83$ to
$1.18$ over parallel CPU. The average CPU utilization of GPU, single
core CPU, and parallel CPU is $64\%$, $82\%$, and $598\%$ respectively. For
Case Study 3, speedup is more significant, with $6.3$ average speedup of
GPU over single core CPU, and $1.75$ over parallel CPU. The
average CPU utilization of GPU, single core CPU, and parallel CPU is $73\%$,
$95\%$, and $680\%$ respectively. Thus, the parallel CPU implementation is showing large speedup similar to the GPU implementation, yet also results in a commensurate CPU utilization percentage, since most cores of the system are fully utilized.

\begin{center}
\vspace{-1mm}
\fbox{\rule{1mm}{0mm}
\begin{minipage}[t]{.9\columnwidth}
{\em Although the parallel CPU implementation provides reasonable speedup, and the single-core CPU implementation imposes low CPU utilization overhead, the GPU implementation manages to achieve both simultaneously.}
\end{minipage}
}\end{center}

%% file: related.tex
\vspace{-4mm}
\section{Related Work}
\label{sec:related}

Runtime verification of parametric properties has been studied by Rosu 
et al~\cite{Hussein,6227231,meredith2013efficient}. In this line of work, it is 
possible to build a runtime monitor parameterized by objects in a Java program. 
The work by Chen and Rosu~\cite{chen2009parametric} presents a method of 
monitoring parametric properties in which a trace is divided into slices, such 
that each monitor operates on its slice. This resembles our method of 
identifying trace subsequences and how they are processed by submonitors. 
However, parametric monitoring does not provide a formalization of applying 
existential and numerically constrained quantifiers over objects.

Bauer et al.~\cite{bauer2013propositional} present a formalization of a 
variant of first order logic combined with LTL. This work is related to our 
work in that it instantiates monitors at run time according to valuations, and 
defines quantification over a finite subset of the quantified domain, normally 
with that subset being defined by the trace. Our work extends this notion with 
numerical constraints over quantifiers, as well as a parallel algorithm for 
monitoring such properties.

The work by Leucker et al. presents a generic approach for monitoring 
modulo theories~\cite{deckermonitoring}. This work provides a more expressive 
specification language. Our work enforces a canonical syntax which is not 
required in~\cite{deckermonitoring}, resulting in more expressiveness. However, 
the monitoring solution provided requires SMT solving at run time. This 
may induce substantial overhead as opposed to the lightweight parallel 
algorithm presented in this paper, especially since it is designed to allow 
offloading the workload on GPU. SMT solving also runs the risk of 
undecidability, which is not clear whether it is accounted for or not. \LTLFOC 
is based on six-valued semantics, extending \LTLfour by two truth values: $\TC$ 
and $\FC$. These truth values are added to support quantifiers and their 
numerical constraints. This six-valued semantics provides a more accurate 
assessment of the satisfaction of the property based on finite traces as opposed 
to the three-valued semantics in~\cite{deckermonitoring}. Finally, although 
\LTLFOC does not support the expressiveness of full first-order logic, 
numerical constraints add a flavor of second-order logic increasing its 
expressiveness in the domain of properties where some percentage or count of 
satisfied instances needs to be enforced.

The work in~\cite{barre2013mapreduce} presents a method of using MapReduce to evaluate LTL properties. The algorithm is capable of processing arbitrary fragments of the trace in parallel. Similarly, the work in~\cite{basinscalable} presents a MapReduce method for offline verification of LTL properties with first-order quantifiers. Our work uses a similar approach in leveraging MapReduce, yet also adds the expressiveness of counting semantics with numerical constraints. Also, our approach supports both offline and online monitoring by introducing six-valued semantics, which are capable of reasoning about the satisfaction of a partial trace. This is unclear in~\cite{basinscalable}, since there is no evidence of supporting online monitoring.

Finally, the work in~\cite{bbf13} presents two parallel algorithms for 
verification of propositional \LTL specifications at run time. These algorithms 
are implemented in the tool RiTHM~\cite{njsbmfb13}. This paper enhances the 
framework in \cite{bbf13, njsbmfb13} by introducing a significantly more 
expressive formal specification language along with a parallel runtime 
verification system.

%% file: conclusion.tex
\vspace{-2mm}
\section{Conclusion}
\label{sec:concl}

In this paper, we proposed a specification language \linebreak (\LTLFOC) for 
runtime verification of properties of types of objects in software and 
networked systems. Our language is an extension of 
\LTL that adds counting semantics with numerical constraints. The six truth 
values of 
the semantics of \LTLFOC allows system designers to obtain informative 
verdicts about the status of system properties at run time. We also introduced 
an efficient and effective parallel algorithm with two implementations on 
multi-core CPU and GPU technologies. The results of our experiments on three 
real-world case studies show that runtime monitoring using GPU provides us with 
the best throughput and CPU utilization, resulting in minimal intervention in 
the normal operation of the system under inspection.

For future work, we are planning to design a framework for monitoring \LTLFOC 
properties in distributed systems and cloud services. Another direction is to 
extend \LTLFOC such that it allows non-canonical strings of quantifiers. 
Finally, we are currently integrating \LTLFOC in our tool 
RiTHM~\cite{njsbmfb13}.